\documentclass{kapproc} 

\usepackage[dvips]{graphicx}

\upperandlowercase

\kluwerbib 

\begin{document}

\articletitle[Pseudobulges in Barred S0 Galaxies]
{Pseudobulges in Barred S0 Galaxies}

\author{Peter Erwin\altaffilmark{1}, John E. Beckman\altaffilmark{1},
and Juan Carlos Vega Beltr\'an\altaffilmark{1}}
 
\affil{\altaffilmark{1}Instituto de Astrof\'{\i}sica de Canarias, La 
Laguna, Tenerife, Spain}

\begin{abstract}
We present preliminary results from an ongoing study of the bulges of
S0 galaxies.  We show that in a subsample of 14 \textit{barred} S0
galaxies, fully half the photometrically defined bulges show kinematic
signatures of \textit{pseudobulges} -- that is, their kinematics are
dominated by rotation.  In four of these galaxies, we identify at
least two subcomponents in the photometric bulge region: flatter,
disk or bar components, assocated with disklike kinematics, and
rounder ``inner bulges,'' which appear to be hotter systems more like
classical bulges.
\end{abstract}


\section{Introduction}

The centers of disk galaxies are conventionally thought to be
dominated by a ``classical'' bulge: a spheroidal, kinematically hot
structure similar to a small elliptcal galaxy, which produces the
central excess of light above that of the exponential disk.  However,
Kormendy (1982, 1993) argued that in at least some galaxies the
central light came from flattened, disklike components he termed
``pseudobulges,'' possibly the result of bar-driven gas inflow and
star formation.  At present, very little is known about the
demographics of such structures.  To address this, we are undertaking
a systematic study of bulge morphology and kinematics in early-type
disk galaxies.


\section[A High Fraction of S0 Pseudobulges]
{A High Fraction of S0 Pseudobulges}

We have performed a preliminary analysis for about two-thirds of the
barred S0 galaxies in our sample, using long-slit spectroscopy from
the ISIS spectrograph of the 4.2m WHT and kinematic data from the
literature (Kormendy 1982; Simien \& Prugniel 2000; Caon et al.\
2000).  The results are shown in Figure~1, which plots
$V_{\rm{max}}/\sigma$ for the photometric bulge regions of fourteen
barred S0 galaxies.  Half of the galaxies lie on or slightly below the
isotropic oblate rotator line, indicating kinematics like those of
classical bulges or less luminous ellipticals (e.g., Davies et al.\
1983).  But the other half lie above, making them kinematic
pseudobulges.  The seven pseudobulges see here imply a lower limit of
30\% for the entire sample of barred S0s.  For four of these seven --
NGC~2787, NGC~2950, NGC~3945, and NGC~4371 -- we have morphological
and/or kinematic evidence that the pseudobulge consists of
\textit{two} components (e.g., Erwin et al.\ 2003): a disklike region
(distinct from the galaxy's main, outer disk) and a central region
more like a classical bulge.

\begin{figure}[ht]
\vskip.2in
\centerline{\includegraphics[width=3.8in]{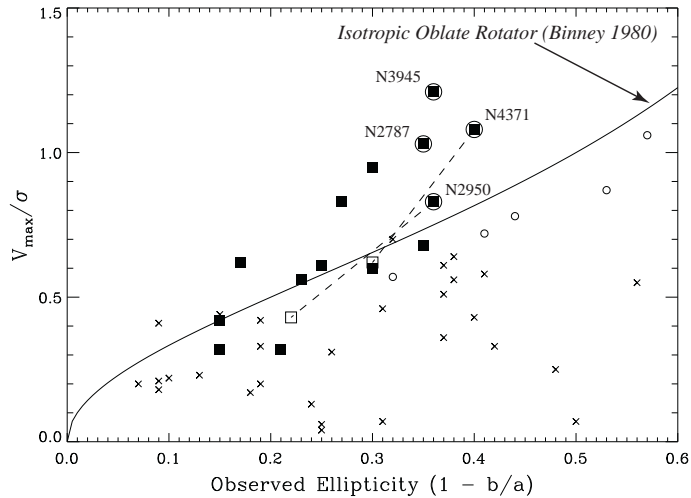}}
\caption{Stellar kinematics of barred S0 bulges: the ratio of maximum
velocity $V_{\rm{max}}$ to mean velocity dispersion $\sigma$ in the
photometric bulge region versus the ellipticity of that region.  The
solid line is the relation for isotropic oblate rotators (i.e.,
``classical'' bulges and lower-luminosity ellipticals; Binney 1980);
galaxies below it are probably dominated by velocity anisotropy, while
galaxies above it are dominated by rotation.  The crosses are
elliptical galaxies from Davies et al.\ (1983); the open circles are
edge-on bulges in early-type disk galaxies (Kormendy \& Illingworth
1982; Jarvis \& Freeman (1985).  The barred S0 bulges from our sample
are filled boxes.  Circled bulges have morphological evidence for
multiple components in the photometric bulge region; open boxes
indicate kinematics for the innermost regions of NGC~2950 and
NGC~4371.}
\end{figure}

\end{document}